\documentclass[twocolumn,twoside,slac_two]{revtex4}
\usepackage{graphicx}
\usepackage{fancyhdr}
\begin{document}

\title{Emerging Computing Technologies in High Energy Physics}

%

\author{Amir Farbin}
\affiliation{Department of Physics, University of Texas at Arlington, Arlington, TX 76019}

\begin{abstract}
  While in the early 90s High Energy Physics (HEP) lead the computing
  industry by establishing the HTTP protocol and the first
  web-servers, the long time-scale for planning and building modern
  HEP experiments has resulted in a generally slow adoption of
  emerging computing technologies which rapidly become commonplace in
  business and other scientific fields. I will overview some of the
  fundamental computing problems in HEP computing and then present the
  current state and future potential of employing new computing
  technologies in addressing these problems. 
\end{abstract}

\maketitle

\thispagestyle{fancy}


\section{HEP Computing Problems}

The high energy frontier will soon be explored by four detector experiments recording the results of collisions of the Large Hadron Collider (LHC)located at the European Center for Nuclear Research (CERN) in Geneva, Switzerland. Designed to record, identify, and study the Higgs boson and a wide array of potential new physics signatures as well as $B$ mesons and quark-gluon plasma,  these experiments ultimately hope to observe an inconsistency between nature and the Standard Model (SM) of High Energy Physics. However, since the typical production and detection probabilities for interesting processes are eleven orders of magnitude smaller than that of uninteresting SM backgrounds, these experiments must sift through forty million events per second while the LHC is colliding beams, recording roughly one in two hundred thousand events in order to accommodate bandwidth constraints. The resultant 1 to 2 petabytes (PB) of monthly data must then be processed and analyzed offline into a form which is suitable for extraction of measurements.

The unprecedented scale of the required computing resources and the complexity of the computing challenges have made computing an important element of HEP. Each LHC experiment has deployed its own ``Computing Model'' (CM), which consists of several classes (``tiers'') of facilities which together comprise a grid of resources held bound by fast links and special middleware.  In ATLAS experiment, the Tier-0 facility at CERN will perform the first-pass processing of the data. The ten national Tier-1 facilities will reprocess these data with better calibrations within two months after data collection. Meanwhile, the roughly thirty Tier-2 facilities placed at specific universities and labs will focus on simulations and data analysis. In addition, considerable effort is directed towards development of:

\begin{itemize}
\item Monte Carlo tools (which link theory and experiment),
\item detector simulation frameworks,
\item algorithmic and statistical analysis tools,
\item data processing frameworks and algorithms,
\item grid middleware,
\item data production and management systems, and
\item underlying data persistency and database infrastructure.
\end{itemize}
Ultimately the performance of these software components, many of which are used by multiple experiments, determines the scale of computing resources required by HEP experiments.

\begin{figure}[h] \centering \includegraphics[width=80mm]{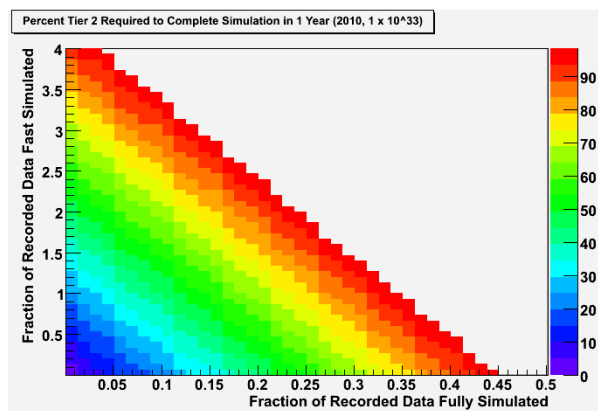} \caption{Percentage of ATLAS Tier 2 CPU required for simulation production as function of fraction of 2010 recorded data which is fully and fast simulated.} \label{fig:CM} \end{figure}

Despite the impressive scale of the LHC computing grid and the sophistication of underlying software technologies, a dearth of computing resources will be one of the primary bottlenecks in extracting measurements from LHC data. For example, 
figure~\ref{fig:CM} shows the percentage of ATLAS Tier 2 resources required in 2010 for fast and full simulation production as function of the fraction of recorded data. The ATLAS CM model expects that physics analysis activity will require roughly half of the resources at ATLAS Tier 2s, leaving the other half for simulation. But we see that the allocating the nominal 50\% of Tier 2 resources to simulation limits the volume of fully simulated data to roughly 20\% of the data the ATLAS detector will collect. This deficiency will fundamentally limit the significance of the comparisons between recorded data and theoretical predictions (via detailed detector modeling) which are necessary to make any statements about nature. Therefore ATLAS must rely on less accurate fast simulations to produce Monte Carlo statistics comparable to the recorded data. And if data needs to be resimulated, as is often the case in the first years of an experiment, CPU resources must be borrowed from analysis activity, thereby stalling extraction of measurements. 

In addition, the experiment CMs do not typically provide physicists the necessary resources for the CPU-intensive activities which in the past decade have come to characterize analyses of HEP data at the Tevatron and the $B$ factories. These activities include sophisticated fits, statistical analysis of large ``toy'' Monte Carlo models, matrix element calculations, and use of the latest discriminant techniques, such as boosted decision trees. Physicists must therefore rely on leveraged resources and emerging technologies to accomplish such tasks.

\section{Solutions}

In tandem with these developing needs in high-energy physics, the role of computing in both business and everyday life has evolved significantly. In the early 1990s, the HTML protocol was developed in the context allowing the hundreds of collaborating physicists in HEP experiments at CERN to communicate with their global group of colleagues.  This development was crucial to the later transformation of the Internet from an academic tool to a global medium.  Today, computing is seen as a commodity, and a critical component of the world economy. Major companies like Google and Amazon manage huge data centers and sell CPU cycles, disk, and bandwidth by the minute and megabyte. Regular innovations in data processing, delivery, organization, and communication continuously drive significant business and social progress.

\begin{figure}[h] \centering \includegraphics[width=80mm]{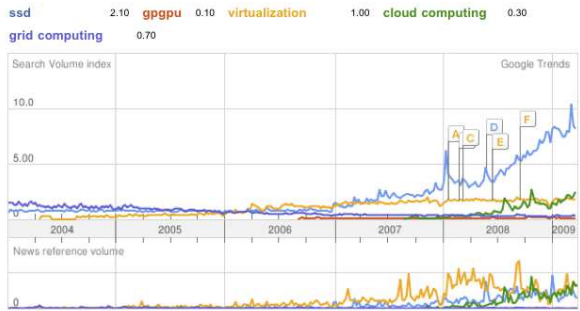} \caption{Search google volume for ``ssd'', ``gpgpu'', ``virtualization'', ``cloud computing'', and ``grid computing'' normalized to ``virtualization''. From Google trends.} \label{fig:googleTrends} \end{figure}

Figure~\ref{fig:googleTrends} compares the Google search volume for ``grid computing'' with technologies such as solid-state drives (SSDs), General Purpose Graphics Processing Units (GPGPU), Virtualization, and Cloud computing. We see that while HEP has been building large and expensive grid sites along with the necessary grid software, Virtualization and Cloud Computing have developed a wider appeal. Fortunately, efforts to take advantage of these technologies have recently begun in HEP.

Virtualization is likely to help address problems in HEP computing that can be traced back to the fact that HEP experiment software generally require specific operating systems (OS) and are typically difficult to install. Virtualization provides a means of providing single file (ie a virtual hard drive image) which can be pre-packaged with all necessary software, including the OS, and can run on any modern ``host'' machine regardless of the host hardware and OS. The first benefit of virtualization is that average physicists can simply download such an file and then instantly be able to use their experiment's software on their personal computer without a great deal of expertise. The CernVM~\cite{cite:CernVM} effort have been particularily successful in this area. Similarly system administrators can use virtualization to simplify deployment of the complicated set of software packages required by GRID sites. This route is particularly attractive for Tier 3 sites that do not necessary have a full-time system administrator.  But perhaps the most promissing application of virtualization is in the area of oppertunistic computing where idle desktop computers (for example, in university offices, class-rooms, and labs), can be used to assist data simluation. 

The appeal of Cloud Computing is the promise of on-demand access to vast computing resources hosted by companies that can provide attractive pricing due to the economy of scale. So, for example, a HEP experiment can short-term lease huge CPU resources for data simulation. It is noteworthy that Cloud Computing has an implicit reliance on Virtualization for delivering the appropriate software environment to the cloud resource. At this point, it is not clear that such Cloud Computing is appropriate for HEP. Current Cloud Computing offerings such as Amazon's EC2 are targeted for web-hosting rather than large data processing tasks and therefore offer prohibitive pricing and performance for HEP applications. What's more, the large Cloud Computing providers such as Amazon and Microsoft favor their own proprietry software solutions that lock their clients to their Cloud and have avoided open efforts such the ``Open Cloud Manifesto'' by IBM. In addition some in the industry view the Clound Computing vision as an unrealistic ``myth''. Nonetheless, many within the HEP community are exploring the potential of HEP computing. For example Nimbus now provides a means of turning Amazon EC2 resources into a self-configuring cluster for HEP computing.

A very promising technology, that is now becoming cost-effective, is SSD. Because these storage devices have no mechanical or moving parts, they provide impressive input/output (I/O) rates, in particular huge gains in random read access in comparison to traditional hard drives (HD). Two high I/O HEP applications that can benefit from SSDs immediately come to mind. First is ntuple data analysis, a task that is typically characterized by rapid iterations over upto terabytes of data. The second is in computing sites where hundreds of processes running on multiple cores access data stored on a single storage device. Since SSDs present the same interface as HDs, their deployment in HEP environments is rather easy. The primarly limitation is then the additional cost of the SSDs, which we may expect to lead to SSD/HD hybrid solutions where HDs are used for long-term storage and data is prestaged to SSDs for faster access. 

We may also note that while HEP has traditionally interpreted Moore's law as predicting faster processors, the industry has shifted to more cores per processor. Thus we find that though many HEP applications in principle lend themselves to parallelization, very few existing HEP software were fundamentally designed with parallel processing on multiple cores in mind.  Recently, the omnipresence of dual and quad-core processors has prompted some parallelization efforts within the HEP community. These efforts take advantage of the embarrassingly parallel nature of HEP computations by simply running multiple instances (or threads) of the software. The only challenge then is sharing memory across these parallel threads in applications that have a large memory footprint such as simulation and reconstruction. Assuming that memory cost isn't a factor, the problem then becomes the processor to memory bandwidth which can be saturated when a large number of cores access large amount of memory. This bandwidth limitation constrains performance gain when scaling to large number of cores per processor. As a result, most HEP multi-core optimization efforts are concentrated on processing forking and thread safety.

While Central Processing Units (CPUs) have been evolving towards more cores, Graphics Processing Units (GPUs) that were originally targeted to the personal computer gaming enthusiasts have evolved to be capable of computing traditionally performed on the CPU. These modern processors that are also now present in many desktops and computers are known as the General Purpose Graphical Processor Units (GPGPUs). Despite that fact that for specific computing tasks, these GPGPUs can marshall thousands of simplified processing units in parallel in order to reduce computing times by orders of magnitude,  they have yet to capture the attention of HEP as a whole. 

\section{Simple ROOT Analysis}

A very simple study of read/write rates of ROOT~\cite{cite:ROOT} analyses illustrates the potential of SSDs and GPGPUs on the data analysis iteration rate. For this study, we consider two simple ROOT applications, one which creates an ntuple (TTree) with random  data (simple types such as bools, ints, floats, and vectors of simple types), and another that reads histograms all quantities in these ntuples. We find that simple read/write rates stablilize with about 20 variables of each type per event (3 KB/event) and 600 events. Table~\ref{fig:SSDHD} summarizes the results of the study. With ROOT's data compression turned off, a single instance of each application achieve approximately 25 MB/s read or write rate on a hard drive which provide 70 MB/sec sequential read. With compression turned on (providing 30\% file size reduction), this figure falls to 4 and 16 MB/s read and write, respectively, illustrating that (de)compression is the main bottleneck in input/output bound ROOT analyses. Ideally, the GPU can be used to eliminate this bottleneck. In order to observe the benefits of the fast random access of SSDs, we ran eight instances of these applications with the data stored on a single HD or SSD. We also observe that only uncompressed data writing appears to be limited by disk access. And though SSDs generally provide faster rates, the improvement is significantly less where the I/O is limited by (de)compression.

\begin{table*}[t]
\begin{center}
\caption{Read/write rates for example ROOT analysis as function of number of simultaneous instances of analysis and ROOT compression level.}
{ 
\begin{tabular}{|l|c|c|c|c|}
 \hline 
Task & \textbf{ Simultanous} & \textbf{ Compression} & \textbf{ HD Rate} & \textbf{ SSD Rate} \\
        & \textbf{ Jobs}               & \textbf{ Level}              & \textbf{ (MB/s)}    & \textbf{ (MB/s)} \\
\hline Writing & 1 & 0 & 25 & 25 \\
\hline Writing & 8 & 0 & 50 & 75 \\
\hline Writing & 1 & 9 & 4   & 25 \\
\hline Writing & 8 & 9 & 25 & 32 \\
\hline Reading & 1 & 0 & 24 & 32 \\
\hline Reading & 8 & 0 & 20 & 28 \\
\hline Reading & 1 & 9 & 16 & 20 \\
\hline Reading & 8 & 9 & 14 & 17 \\
\hline
\end{tabular}
}
\label{fig:SSDHD}
\end{center}
\end{table*}

\section{GPGPUs}


From the Cell processor in the Playstation 3 (PS3) to in newer-generation Graphic Processing Units (GPUs) used in desktop and laptop computers, we find the building blocks for High-Performance Computing (HPC) systems already present in devices we daily use. Originally driven by the gaming industry, GPU architectures have recently been developed to also support general-purpose computing. These GPUs offer impressive power consumption/performance and price/performance ratios. The omnipresence of these general purpose GPUs have pushed industry leaders such as Microsoft and Apple into a race to develop strategies that take advantage their capabilities. The raw computational horsepower of GPUs is staggering. A single modern GPUs provides nearly one TFLOPS ($10^{12}$ floating-point operations per second)~\cite{Talton:2006cz}, roughly 20 times more than a typical multi-core CPU. What's more, the trend over the past decade exhibits an exponential growth in the ratio of the computational power of GPU to CPU.

While use of computer graphics hardware for general-purpose computation has been an area of active research for many years (eg ~\cite{England:1978vh}, ~\cite{Potmesil:1989sy} \cite{Rhoades:1992bl}), the wide deployment of GPUs in the last several years has resulted in an increase in experimental research with graphics hardware. Some notable work include password cracking~\cite{Kedem:1999jx}, artificial neural networks~\cite{Bohn:1998df}, solving partial differential equations (PDEs)~\cite{Harris:2002to}, line integral convolution and Lagrangian-Eulerian advection~\cite{Heidrich:1999il,Jobard:2001eb,
  Weiskopf:2001xw}, and protein folding
(Folding@Home)~\cite{folding}. In HEP, the developers of the FairRoot
demonstrated a two orders of magnitude acceleration of track fitting
in high multiplicity environments using GPUs~\cite{fairroot}.

Currently, the primary players in the GPU arena are NVIDIA and AMD (through the purchase of ATI). Nearly all of NVIDIA's GPUs, from the low-end laptop GPU to the professional TESLA line, can be programmed using their Compute Unified Device Architecture (CUDA), a parallel programming compiler and software environment designed for issuing and managing general purpose computations on the GPU through extensions to the standard C language. There are also standard numerical libraries for FFT (Fast Fourier Transform) and BLAS (Basic Linear Algebra Subroutines).

ATI/AMD has also developed a proprietary GPGPU hardware and software known as Data Parallel Virtual Machine (DPVM). However, their focus seems to now shifted to Open Computing Language (OpenCL), a GPU software standard originally developed by Apple and then released to an open consortium which includes all the relevant companies as signatories~\cite{OpenCL}. The primary appeal of OpenCL is it's architecture independence which is achieved through run-time compilation of computing software kernels. The first publicly available implementation of OpenCL was recently released last week as part of Apple's Snow Leopard operating system.

Other, less popular, GPU architectures and software include ClearSpeed~\cite{ClearSpeed}, BrookGPU
 ~\cite{BrookDocs}, AMD Stream Computing, and Sh~\cite{Sh}. Finally, a very promising upcoming product is Intel's Larabee, a x86-based many-core GPU which is rumored to be release later this year.

\subsection{Developing GPGPUs Applications}
The processing model for GPU is very different than CPU. Whereas CPUs are optimized for low latency, GPUs are optimized for high throughput.  GPUs are essentially stream processors, hardware that operates in parallel by running a single computation on many records in a stream at once. The limitation is that each parallel computation may only process independent memory elements that cannot share memory with others. The result is that GPUs are generally suitable for computations that exhibit specific characteristics. They must be compute intensive with large number of arithmetic operations per I/O or global memory reference.  They must be amiable to data parallelism, where the same function is applied to all records of an input stream and a number of records can be processed simultaneously without waiting for results from previous records. And they must exhibit data locality, a specific type of temporal locality common in signal and media processing applications, where data is produced once, read once or twice later in the application, and never read again.

HEP applications that are good candidates for GPGPU acceleration include Monte Carlo integration for matrix element methods, discriminant training or calculation in multivariate analysis, maximum-likelihood fitting, compression/decompression during data input/output, event generation, full or fast detector simulation, event reconstruction (in particular for the High Level Trigger), and detector alignment. The difficulty with employing GPGPUs is that existing applications cannot be simply rebuilt to run on GPGPUs, but must rather treat the GPU as a co-processor. The software must explicitly adopt a data-parallel processing model where the data is broken into chunks and independently processed by algorithms which are highly constrained in both their memory access and complexity.  A practical approach of GPGPU accelerating existing applications is to rewrite computational bottlenecks to prepare the data on the CPU, transfer it to the GPU memory, execute the computation on the GPU, and then transfer the results back.

Even modest acceleration of detector simulation and reconstruction using GPGPU will have a significant impact on HEP computing. For reconstruction, where thousands of tracks and hundreds of thousands calorimeter cells must be processed, any gain directly translates to higher trigger output. For simulation, where thousands of particles must be propagated through detector and magnetic fields, gains translate into fewer computing resource requirements. The practical time-scale for deployment of such GPGPU accelerated strategies is for LHC upgrade. While strategies for GPGPU acceleration of tracking and calorimetry are rather straight-forward, the complexity of Geant4~\cite{Agostinelli:2002hh} and the fact that it wasn't written with parallelization in mind, make simulation a much more difficult problem. Taking full advantage of GPGPUs in simulation will likely only be possible in the next generation software (perhaps Geant5). One promising interim strategy is to employ multiple parallel Geant4 threads running on the CPU (see Geant4 parallelization efforts of \cite{cite:northeastern}) that offload specific calculations (eg magnetic field extrapolation) to a service that can batch perform the calculation using the GPU.

\end{document}